\documentclass[preprint,showpacs,amsmath,amssymb]{revtex4}

\usepackage{amsmath}
\usepackage{graphicx}
\usepackage{epsfig}

\begin{document}

\title{Density Functional Theory of Multicomponent Quantum Dots}

\author{K. K\"arkk\"ainen$^1$, M. Koskinen$^1$, S.M. Reimann$^2$, and M. Manninen$^1$}
\affiliation{$^1$Department of Physics, University of Jyv\"askyl\"a,
FIN-40351 Jyv\"askyl\"a, Finland}
\affiliation{$^2$Mathematical Physics, Lund Institute of Technology, Lund, Sweden}

\date{\today}

\begin{abstract}
Quantum dots with conduction electrons or holes originating from several bands are considered.
We assume the particles are confined in a harmonic potential and assume the electrons
(or holes) belonging to different bands to be different types of fermions with 
isotropic effective masses. The density functional method with the local density approximation
is used.
The increased number of internal (Kohn-Sham) states leads to a generalisation of 
Hund's first rule at high densities. At low densitites the formation of Wigner molecules 
is favored by the increased internal freedom.     
\end{abstract}

\maketitle

\section{Introduction}

In simple models of quantum dots, the conduction electrons (or holes) of a
semiconductor are confined into a two-dimensional harmonic 
trap (for reviews see \cite{chakraborty1999,reimann2002}). The band 
structure of the material is taken into account through the effective mass approximation,
and screening effects are accounted for by the dielectric constant.
The problem is then reduced to solving the many-particle problem of interacting 
electrons in a two-dimensional harmonic potential. The electrons have 
spin as an 'internal degree of freedom'. Neglecting spin-orbit coupling, 
the spin-up and spin-down electrons can be treated as separate interacting particles,
and consequently we can say that the normal electron gas is a two-component
gas, the components being the spin-up and spin-down electrons.
Similarly, we will call the polarised electron gas as a 
one-component system (sometimes also called a system of spin-less fermions).

The simple picture will fail in describing more complex structures 
where the number of degrees of freedom of the electrons is increased 
either by several two-dimensional layers or by multiple valleys of the band
structure. For example, in a vertical double-layer quantum dot
the electrons confined in the two layers form (in the vertical direction)
''odd'' and ''even'' states \cite{partoens2000,pi2001,pi2001b}
which could be approximated as different components,
or as different isospin-states of the electron \cite{hawrylak1995}. The isospin together
with the spin would make the system a four-component electron gas.

Other multicomponent electron systems would be quantum
dots in silicon. In this case the conduction electrons originating from 
four valleys of the conduction band 
could be approximated as different, but mutually interacting fermions.
Similarly, quantum dots with holes would always have particles belonging to 
different bands, i.e. heavy holes and light holes \cite{marder2000}.

Density functional theory in the local (spin) density approximation (LSDA) 
provides a flexible method 
to study the ground state properties of interacting electrons in quantum dots 
\cite{reimann2002}. In LSDA the 
exchange and correlation effects of the interacting conduction electrons 
are locally approximated by the exchange-correlation energy ($\epsilon_{xc}$) 
of the two-dimensional, homogenous gas.
Similarly, in a multicomponent electron system the starting point for the local
density approximation is the exchange-correlation energy of a two-dimensional
multicomponent gas. Recently, we have suggested that 
the multicomponent $\epsilon_{xc}$  can well be approximated by 
extending the parametrised two-component function 
of Attaccalite \emph{et al.} \cite{attaccalite2002} to a higher 
number of components \cite{karkkainen2003}. 
This parametrisation was shown to have correct high and low density
behaviour and it has rather simple dependence on the densities of different components. 
Moreover, the dependence of the effective masses of the components
can be approximated through a simple scaling of the density parameter by
an effective average mass. 

In this paper we study the electronic structure of quantum dots where
the electron gas has from one up to eight internal degrees of freedom,
using the exchange-correlation functional suggested in Ref. \cite{karkkainen2003}.
The inclusion of more components increases the number of 
internal degrees of freedom in the system. As a result, in the self-consistent scheme 
the electrons can access more Kohn-Sham states in addition to the usual two spin states, 
leading to new features in the electronic structure.
We will discuss the electronic shell structure of four-component quantum dots by 
studying the addition energy spectrum at high densities. 
In the low density limit the formation of Wigner molecules in an eight-component 
quantum dot is investigated. We found that the seven-electron configuration is 
particularly stable like the classical point-charge calculation predicts \cite{bedanov1994}. 
In the Wigner molecule limit the fermions of different components are distributed 
spatially so that antiferromagnetic frustration is avoided. Finally we notice that even a slight
increase of mass favors the heavier components as the heavy component states are pushed down
in energy.

\section{theoretical model}

We write the total density of the multicomponent electron gas as
\begin{equation}
n({\bf r})=\sum_{i=1}^{\Lambda}n_i({\bf r})=n({\bf r})\sum_{i=1}^{\Lambda}\nu_i({\bf r}),
\end{equation}
where $\Lambda$ is the number of components, 
$n_i$ are the densities of components and $\nu_i$ dimensionless concentrations
of the components. In the multicomponent case, the use of concentrations is simpler than
the use of total density and polarization $\zeta$, as usually done for the normal electron gas
(in the two-component case, $n_1=n_\uparrow$ and $n_2=n_\downarrow$ and
$\zeta=\nu_1-\nu_2$). Following the notations of Ref. \cite{karkkainen2003} we define
numbers
\begin{equation}
Z_\gamma=\sum_{i=1}^\Lambda \nu_i^\gamma,
\label{znumber}
\end{equation}
which allows us to write the exchange-correlation functional of Attaccalite 
{\it et al}. \cite{attaccalite2002} as 
\begin{equation} \label{eq:exc}
\epsilon_{xc}(r_s,\{\nu_i\})=e^{-\beta r_s}[\epsilon_x-\epsilon_x^{(6)}]+\epsilon_x^{(6)}+\alpha_0(r_s)
	+\alpha_1(r_s)(2Z_2-1)+\alpha_2(r_s)(2Z_2-1)^2~,
\end{equation}
where  $\epsilon_x^{(6)}=(1+\frac{3}{8}(2Z_2-1)+\frac{3}{128}(2Z_2-1)^2)\epsilon_x(r_s,\zeta=0)$.
As shown earlier \cite{karkkainen2003} this analytic continuation of the originally
two-component functional approximates very well all existing results for the 
exchange-correlation energy of a multicomponent electron gas.

The exchange energy is independent of the particle mass. If the effective masses of 
all the particles are the same, the mass dependence of the total energy becomes 
just a scaling factor. If the masses are not the same we can use as a first
approximation a properly weighed average mass as a scaling factor and write \cite{karkkainen2003}
\begin{equation}
\epsilon_{xc}(r_s,\{ \nu_i \},\{ m_i \})=\frac{M}{m_e}\epsilon_{xc}(Mr_s,\{ \nu_i\},\{ m_i=m_e\})~,
\label{scale}
\end{equation}
where $m_e$ is the bare mass of the electron (or any suitably chosen effective mass),
$m_i$ is the effective mass of component $i$, and $M$ is an average mass defined as
\begin{equation}
\frac{1}{M}=\frac{1}{Z_2}\sum_{i}^{\Lambda}\frac{\nu_i^2}{m_i}.
\end{equation}
Note that the density parameter $r_s$ in our formulation always refers to the total
number density of all particles: $r_s=1/\sqrt{\pi n}$.

The Kohn-Sham equations have to be solved self-consistently for each
component of the electron gas. The effective potential $V_{{\rm eff},i}$ consists of the 
external harmonic confinement (assumed here to be the same for all components),
of the Coulomb repulsion of the electron density distribution (Hartree term) and 
of the exchange-correlation potential, which can be directly derived from the 
multicomponent exchange-correlation energy:
\begin{equation}
V_{{\rm eff},i}({\bf r})=\frac{1}{2}Kr^2+\int d{\bf r}'
\frac{{e^2}n({\bf r}')}{4\pi\varepsilon_0\epsilon\vert{\bf r}-{\bf r}'\vert}+
V_{{\rm exc},i}(r_s({\bf r}),\{\nu_i({\bf r})\},\{m_i\}),
\end{equation}
where $K$ is the strength of the external confinement, $\epsilon$ the dielectric constant.
Note that the exchange-correlation potential depends locally on the 
concentrations of each component.
The Kohn-Sham equations to be solved simultaneously for all components are
\begin{equation}
-\frac{\hbar}{2m_i}\nabla^2\psi_{i,k}({\bf r})+V_{{\rm eff},i}({\bf r})\psi_{i,k}({\bf r})
=\epsilon_{i,k}\psi_{i,k}({\bf r}),
\end{equation}
resulting in the densities $n_i$ of all components $i$,
\begin{equation}
n_i({\bf r})=\sum_k^{N_i}\left\vert\psi_{i,k}({\bf r})\right\vert^2,
\end{equation}
where $N_i$ is the number of electrons of component $i$. In the ground state
the lowest single particle levels are filled and the numbers $N_i$ are known 
only after the ground state is found. In practice we solve the equations
by keeping the numbers $N_i$ fixed and then choose the configuration
which gives lowest total energy.

The Kohn-Sham equations were solved using a plane-wave expansion
and the fast Fourier transform technique. We use up to $23\times 23$ 
plane waves and, correspondingly a $45\times 45$ lattice at which the 
density and potential is derived. 
When iterating the Kohn-Sham equations, 
a mixing of the new and old potential is necessary to obtain convergence. 
The convergence is slow especially at low densities where broken
symmetry solutions (localization of electrons to Wigner molecules) emerge.
We use effective atomic units where energy is given in 
effective Hartree, Ha$^*$=$m_e^*e^4/\hbar^3(4\pi\epsilon_0\epsilon)^2$ 
and the unit of distance is the effective Bohr radius 
$a_0^*=\hbar^2 4\pi\epsilon_0\epsilon/m_e^*e^2$.

\section{Results}


\subsection{Shell structure -- Addition spectrum}

As confined, fermionic quantum systems, quantum dots show shell structure with degeneracies 
determined by the symmetries of the confining potential. The independent electron
energy spectrum for a 2D harmonic oscillator is $\varepsilon_{nl}=\hbar\omega_0(2n+|l|+1)$,
where $n$ is the principal quantum number and $l$ is the angular momentum (or its z-projection).
At high densities the independent electron scheme describes the qualitative characteristics 
of the shell filling of the interacting system reasonably well, as known from the earlier 
studies \cite{reimann2002} of normal quantum dots. The main effect of the electron-electron
interaction is to produce the spin determined by Hund's first rule \cite{koskinen1997}.
The shell structure effects are usually shown in a form of addition energy spectrum where the
second difference of the total energy is plotted as a function of the number of electrons 
in the dot. Addition energies measure the changes
$\mu(N+1)-\mu(N)$ in the chemical potential $\mu(N)=E(N)-E(N-1)$. 

We will first study the addition energy spectrum of an ideal four-component electron system
in a quite high electron density, $r_s=2\ a_0^*$. A physical realization of such system could be 
a vertical double dot, where the interlayer distance is very small.
Figure 1 shows the addition energy spectrum and a schematic picture of the single 
particle levels. Since each single particle level can now occupy four electrons
the closed shells correspond to total electron numbers 4, 12, 24, etc. These
'magic' numbers are seen as pronounced maxima in the addition energy spectrum. 
In between, we see smaller maxima at every even electron number up to 12 electrons
and at every third electron number between 12 and 24. These are manifestations of
Hund's first rule generalized to the multicomponent case: Degenerate states are filled 
one component at a time. This minimizes the total energy, since the 
exchange energy favors a 'polarized' electron gas.
For example, in the third shell we see maxima at $N=15$, 18, and 21,
corresponding to filling the three orbitals of the first, second and third component of 
the four-component electron gas.

\subsection{Formation of Wigner molecules}

At low enough densities the electrons in the parabolic quantum dot are expected to 
form Wigner molecules as the correlation effects start to dominate the electronic structure.
In this limit the electrons localise in the classical configuration that minimises 
the electrostatic repulsion \cite{bedanov1994} and this charge distribution symmetry is 
reflected in the 
internal structure of the many-body wavefunction \cite{reimann2000}. In the multicomponent
systems the localisation is eased up due to the fact that electrons can access more than two
internal states. In the low density limit we study an eight component system in 
fixed external confinement with $K=2\cdot 10^{-4}$ effective atomic units. 
This choise corresponds to densities
that are only slightly higher than those where a polarised (one-component) 
state becomes the ground state for different electron numbers. 
At these low densities the multicomponent local density approximation 
localises the electrons in the classically predicted configurations.
Figure 2 shows the total electron densities for seven, eight and nine 
electrons. In the cases of seven and eight electrons, one electron is in the
center and the rest form a ring around it, while in the case of nine electrons
two electrons form a 'dimer' at the center and seven electrons from a ring around them.
The geometries are in perfect agreement with those of classical 
electrons \cite{bedanov1994}. 

Figure 3 shows the addition energy spectrum of the eight-component quantum dot 
at the low density. The spectrum does not any more show features of shell structure and 
Hund's rule, but shows a small kink at $N=7$, in agreement with the maximum in the addition energy
spectrum of the purely classical system, determined from the results of Ref. \cite{bedanov1994},
and also shown in Fig. 3. Classically, the seven electrons can form a perfect 
hexagon with one electron at the center. The quantum mechanical solution is the same as shown 
in Fig. 2. In the density functional theory the localized electrons are not
point charges as in the classical case. As a 
consequence, the spectrum is smoother than that of the classical result. 

The evolution of the ground state as a function of density parameter $r_s$ is investigated
in an eight electron quantum dot with four components. As discussed earlier, 
at high densities $r_s\lesssim 4\ a_0^*$ the ground state obeys Hund's rule giving 
the configuration (3,3,1,1) for the ground state. As the density is lowered,
the electron structure shows a Wigner
molecule-like state already at $r_s=6.0\ a_0^*$. Six localised electrons are at 
the outer circumference 
with two non-localised in the middle, as shown in Fig. 4. 
The six electrons at the outer radius belong to two components and the 
two electrons in the middle occupy also two components. The electrons in the outer ring
are distributed 
spatially so that the two nearest neigbours of each electron belong to other component.
This means that
the densities of the outer components are rotated by $\pi$ with respect to one another.
In this way the system 
will avoid antiferromagnetic frustration. For $r_s\gtrsim 8\ a_0^*$ the electrons localise into
a classically predicted 
configuration with seven electrons at the outer radius and one in the middle.
The frustration is again
avoided by taking the nearest neighbours for each electron from other components,
as shown in Fig. 4.

\subsection{Mass dependence}

The effect of the varying mass was tested in the four-component system with $r_s=2.0\ a_0^*$ 
and $N=24$. The masses of two components were increased ($m_1=m_2=m$) while the other 
two masses were kept constant ($m_3=m_4=1.0\ m_e$). For $m=1.0\ m_e$ the sd-shell 
is filled giving the
''magic'' configuration (6,6,6,6). The mass increase shifts the orbitals of the lighter
components up in energy relative to heavier components, as shown in Fig.5.  
As a consequence, at $m=1.2\ m_e$ the sd-orbitals of the
light components are empty and the fp-orbitals of the heavier components are occupied 
according to Hund's rule leading to configuration (10,8,3,3). 
Already at $m=1.8\ m_e$ the lighter orbitals have only two electrons and the heavier components 
obey Hund's rule resulting in the configuration (12,10,1,1).

\section{Summary and discussion}

We have studied the general features of quantum dots confining a multicomponent electron gas.
At high densities the exchange energy favors polarization of electrons and the degenerate
energy levels are filled with one component at a time. This leads to an addition energy
spectrum which besides the peaks at full shells also shows peaks coming from the 
generalization of Hund's rule.

The increased number of internal degrees of freedom of the electrons make it easier
for the electrons to localize to Wigner molecules. We have demonstrated this for
four- and eight-component systems. In the low density limit the addition energy spectrum then 
does not any more show the electronic shell structure but the geometrical shell structure of
Wigner molecules.

If the different electrons (or holes) have different masses, the localization
pushes the energy states of the heavy particles down as compared to those of the light particles.
If the mass difference is small the addition energy spectrum is expected to be complicated due 
to the mixture of light and heavy particle states. However, if the mass ratio is at a 
typical value of heavy and light holes, say five, the light holes do not play any role
until the dot has several tens of particles.

\begin{figure}[h]
\centerline{\epsfxsize=5in\epsfbox{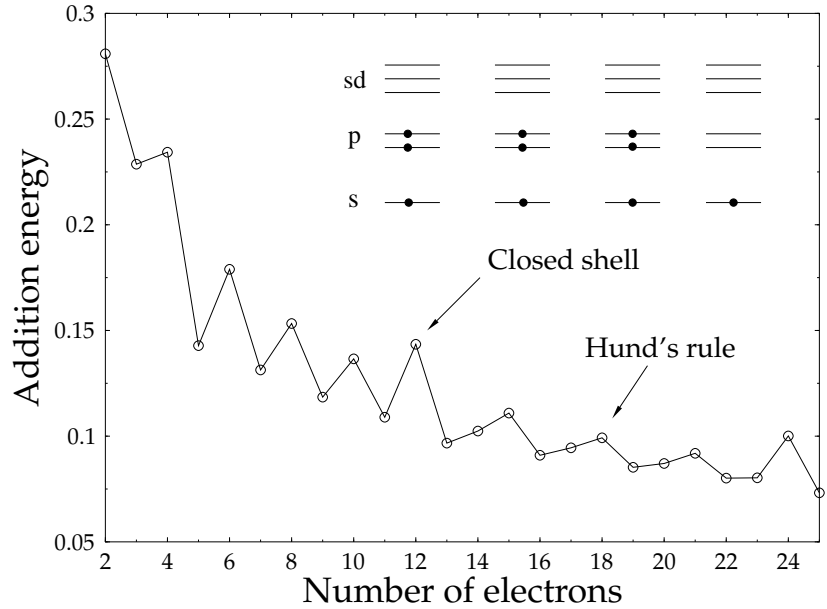}}
\caption{Addition energy spectrum of a four-component quantum dot.
The confinement potential changes with the number of electrons so that the average
electron density in the dot center corresponds to $r_s=2\ a_0^*$. The inset shows schematically the
filling of levels in the case of 10 electrons. The peaks at 4, 12 and 24 are caused
by shell closings while those in between are caused by Hund's rule}
\end{figure}

\begin{figure}[h]
\centerline{\epsfxsize=5in\epsfbox{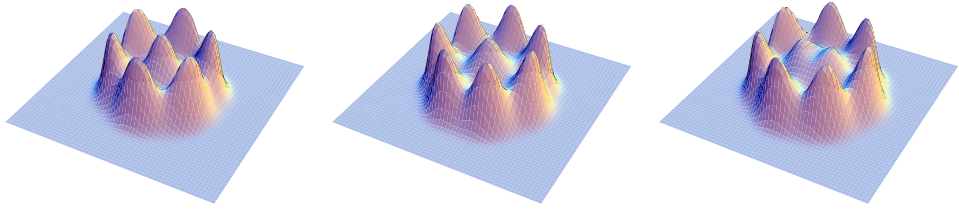}}
\caption{The total electron densities of a quantum dot having
(from left to right) 7, 8 and 9 electrons at low densities. 
The confinement strenght is $K=2\cdot 10^{-4}$ atomic units.
The localization in the local density approximation is made possible by the eight
internal degrees of freedom of the system.} 
\end{figure}

\begin{figure}[h]
\centerline{\epsfxsize=5in\epsfbox{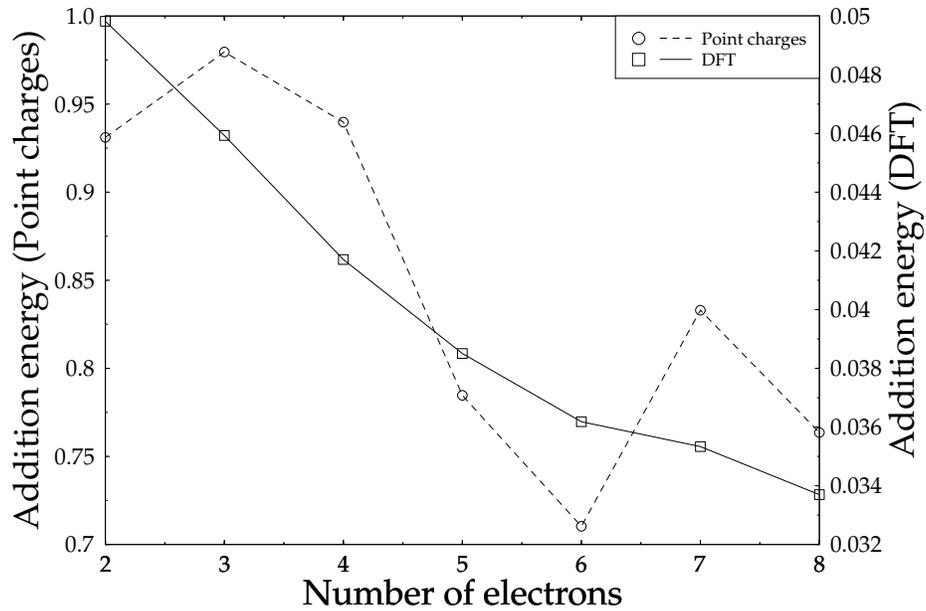}}
\caption{Addition energy for a eight component quantum dot at a low electron density
(the confinement strenght is $K=2\cdot 10^-4$ atomic units).  
The spectrum shows a weak kink at $N=7$ as a precursor of the geometrically
magic structure. 
For comparison, addition energy spectrum of classical electrons is shown.} 
\end{figure}

\begin{figure}[h]
\centerline{\epsfxsize=5in\epsfbox{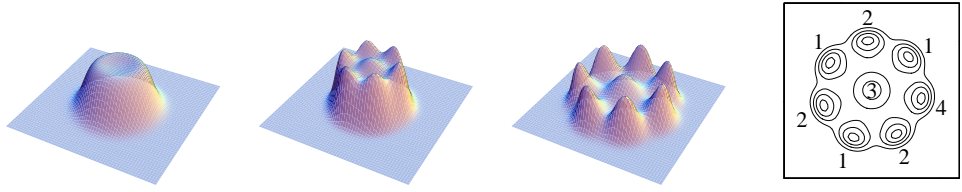}}
\caption{Electron density of a four-cmponent quantum dot for three different
values of $r_s$: From left to right $r_s=2\ a_0^*$, $6\ a_0^*$, and $14\ a_0^*$. 
The localization in the 
multicomponent LDA is made possible by the fact that the neighbouring localized 
electrons belong to different components as indicated by numbers in the contour plot.} 
\end{figure}

\begin{figure}[h]
\centerline{\epsfxsize=5in\epsfbox{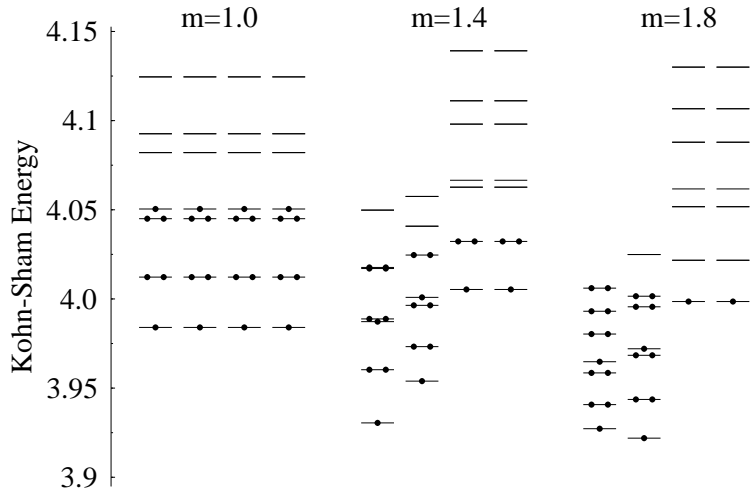}}
\caption{The effect of varying effective mass on the single particle Kohn-Sham levels 
in a four component electron system with 24 electrons. The light components have mass
$m=1\ m_e$ and the heavy mass is indicated in the figure.}  
\end{figure}

\end{document}